\documentstyle[12pt,aps,prb,amsfonts]{revtex}

\tightenlines
\renewcommand{\vec}[1]{{\mathbf #1}}

\begin{document}

\draft

\title{Instabilities of waves in nonlinear disordered media}
\author{S. E. Skipetrov and R. Maynard}
\address{Laboratoire de Physique et Mod\'elisation des Milieux
Condens\'es,\\
Universit\'e Joseph Fourier, Maison des Magist\`eres
--- CNRS,\\
B.P. 166, 38042 Grenoble Cedex 9, France}

\date{\today}
\maketitle

\vspace{1cm}
\begin{abstract}
We develop a self-consistent theory of temporal fluctuations
of a speckle pattern resulting from the multiple scattering
of a coherent wave in a weakly nonlinear disordered medium.
The speckle pattern is shown to become unstable if the nonlinearity
exceeds a threshold value. The instability is due to
a feedback provided by the multiple
scattering and manifests itself in spontaneous fluctuations
of the scattered intensity. The development of instability is
independent of the sign of nonlinearity.
\end{abstract}
\vspace{0.5cm}

Interplay between nonlinearity and multiple scattering has
received considerable attention during the recent years.
Second harmonic generation,\cite{agran88,deboer93}
phase conjugation,\cite{kravtsov90} and
optical Kerr effect\cite{arajuo98}
have been studied in the weak scattering limit
($k \ell \gg 1$, $k$ --- the wavenumber in the medium,
$\ell$ --- the mean free path).
On the other hand, the nonlinearity is shown to modify
the coherent backscattering peak,\cite{agran91}
and to affect the angular correlations
of scattered waves.\cite{bressoux99}

In the present Letter, we consider the {\em temporal\/}
fluctuations of waves in weakly nonlinear disordered media.
We show that, in contrast to linear media, where
the fluctuations of scattered waves
with time are due to scatterer motion,\cite{stephen88,maret87,bicout93}
the wave scattered in a {\em nonlinear\/} medium may exhibit
temporal fluctuations even if the scatterers are immobile.
These fluctuations are {\em spontaneous\/} and represent a signature
of {\em instability\/} of the multiple-scattering speckle pattern.
Although unstable regimes are rather common for nonlinear
waves in feedback systems,\cite{shen84,gibbs85,aless91}
our study is the first to consider
the (multiple) scattering as a feedback-providing mechanism
in a nonlinear medium. The long-range spatial
correlations of intensity fluctuations appear to be crucial
for the feedback efficiency.

We consider a scalar plane wave of frequency $\omega$ incident
upon a nonlinear disordered medium occupying the half-space
$z>0$. Wave amplitude in the medium $\psi(\vec{r}, t)$
obeys a nonlinear wave equation:
\begin{eqnarray}
\left\{ \nabla^2 + k_0^2
\left[ \varepsilon_0 +
\delta \varepsilon(\vec{r}) + \varepsilon_2
\left| \psi(\vec{r}, t) \right|^2 \right] \right\}
\psi(\vec{r}, t) = J(\vec{r}, t),
\label{weq}
\end{eqnarray}
where $k_0$ is the free-space wavenumber,
$\varepsilon_0 = n_0^2$ is the average dielectric function,
$\delta \varepsilon(\vec{r})$ is the
fluctuating part of the dielectric function,
$\varepsilon_2$ is a nonlinear constant, and $J(\vec{r}, t)$
is a (slowly varying)
source term [$\left| (\partial/\partial t) J(\vec{r}, t) \right|
\ll \omega \left| J(\vec{r}, t) \right|$].
Eq.\ (\ref{weq}) describes propagation of waves in
centrosymmetric media, where nonlinear terms quadratic in
$\psi$ are forbidden.\cite{shen84}
Equations of the same type appear for
electromagnetic waves in plasma,\cite{ginzburg64} as well as
in the context of the theory of hydrodynamic turbulence.\cite{zakharov92}
A nonlinear Schr\"{o}dinger equation similar to Eq.\ (\ref{weq})
has been recently considered by Spivak and Zyuzin.\cite{spivak00}

We assume the source term $J(\vec{r}, t)$ in Eq.\ (\ref{weq})
to consist of two parts: $J_0(\vec{r})$, representing an
ideal plane wave incident at $z = 0$, and $\delta J(\vec{r}, t)$,
which is a random noise term. In the present paper, we are only interested
in the case of infinitely weak noise:
$\left| \delta J \right| \rightarrow 0$.
In a linear medium ($\varepsilon_2 = 0$), the solution
of Eq.\ (\ref{weq}) is then unique and time independent:
$\psi(\vec{r}, t) = \psi(\vec{r})$. Consequently, the autocorrelation
function $g_1(\vec{r}, \tau) = \left< \psi(\vec{r}, t) \psi^*(\vec{r}, t+\tau)
\right>/\left< I(\vec{r}, t) \right> \equiv 1$,
where $I(\vec{r}, t) = \left| \psi(\vec{r}, t) \right|^2$
is the wave intensity.
If $\varepsilon_2 \neq 0$, however,
Eq.\ (\ref{weq}) may admit multiple solutions,\cite{spivak00}
and (even being infinitely weak) the noise $\delta J$ can provoke
the system to jump from one solution to another.
Obviously, such
a behavior will result in spontaneous fluctuations of
$\psi(\vec{r}, t)$ with time, yielding $g_1(\vec{r}, \tau) < 1$.
In the general case, the calculation of $g_1$ appears to be
too involved. We therefore limit ourselves to long times
$\tau \gg T_{jump}$, where $T_{jump}$ is the typical time
needed for the system to jump from one solution to another.
Calculation of $T_{jump}$ is beyond the scope of our analysis,
but we expect $T_{jump}$ to be inversely proportional to
the speed of wave in the medium.
For $\tau \gg T_{jump}$, $g_1(\vec{r}, \tau)$
should become independent of $\tau$, and its
asymptotic value can be considered as a measure of the mutual
resemblance of solutions of Eq.\ (\ref{weq}).
To simplify the further analysis,
we adopt the following four assumptions:

\vspace{2mm}
\noindent
({\em i\/}). Propagation of waves in a weakly
nonlinear disordered medium
is diffusive with a mean free path $\ell$ unaffected by nonlinearity.
This implies that nonlinear refraction is
negligible at distances of order
$\ell$, and consequently, that $\Delta n^2 k \ell \ll 1$,
where $\Delta n = n_2 I_0$, $n_2 = \varepsilon_2/(2 n_0)$,
and $I_0 = \left< I(\vec{r}, t) \right>$ in the absence of
absorption.

\vspace{2mm}
\noindent
({\em ii\/}). Statistics of the wave field scattered in
a weakly nonlinear disordered medium is close to Gaussian.
Consequently, the factorization approximation
$g_2(\vec{r}, \tau) =
\left< I(\vec{r}, t) I(\vec{r}, t+\tau)
\right>/\left< I(\vec{r}, t) \right>^2
\simeq 1 + \left|g_1(\vec{r}, \tau)\right|^2$
applies.\cite{bressoux99}
This assumption is justified under
the same conditions as ({\em i\/}), since the Gaussian statistics
of the {\em total\/} scattered wave field $\psi(\vec{r}, t)$
is a consequence of the complete randomization of phases of
{\em partial\/} waves arriving at $\vec{r}$. 
The reason for the randomization is that the typical
distance $\ell$ between individual scattering events in
a multiple-scattering sequence is much larger than the wavelength
($k \ell \gg 1$).\cite{shapiro86}
Obviously, such a mechanism of phase randomization is equally
effective in both linear and weakly nonlinear media,
as long as ({\em i\/}) holds.

\vspace{2mm}
\noindent
({\em iii\/}).
Intensity of the third harmonic
remains always much smaller than the intensity of the
fundamental wave.
This implies that the medium has a sufficient degree of dispersion
for the phase matching condition\cite{shen84} to be violated:
$\left| \Delta k \right| \ell \gg 1$ with
$\Delta k = k_3 - 3 k$ and $k_3$ being the wavenumber at
frequency $3 \omega$.

\vspace{2mm}
\noindent
({\em iv\/}).
Functional form of the $\vec{r}$-dependence
of $g_1$ is the same as in the linear medium:
$g_1(\vec{r}, \tau) = \exp[ - \beta(\tau)\, z/\ell ]$,
where $\beta \ll 1$ is some unknown function.
To justify this ansatz, we make use of ({\em i\/})
and write $g_1$ as a sum of contributions of paths
of various lengths $s$:\cite{maret87,bicout93}
\begin{eqnarray}
g_1(\vec{r}, \tau) = \int_{0}^\infty \mathrm{d} s\,
P(z, s)
\exp\left[ -\frac{1}{2} \left< \Delta \varphi^2(\tau) \right>_s
\right] \approx \int_{0}^{s_1(\tau)} \mathrm{d} s\,
P(z, s).
\label{g1ps}
\end{eqnarray}
Here $P(z, s) \propto z s^{-3/2} \exp[-3 z^2/(4 \ell s)]$
is the relative ``weight'' of paths of length $s$,
and $\left< \Delta \varphi^2(\tau) \right>_s$ is the variance of
the phase difference between two waves propagating along the
same path but separated in time by $\tau$, averaged over all possible
realizations of disorder and over all paths of the same length $s$.
From here on, we denote such an averaging by
$\left< \cdots \right>_s$.
As a function of $s$, $P(z, s)$ is peaked at $s_0 = z^2/(2 \ell)$.
The second equality of Eq.\ (\ref{g1ps}) is obtained by assuming that
$\left< \Delta \varphi^2(\tau) \right>_s$ increases
monotonically with $s$ and becomes of order
unity at $s = s_1(\tau) \gg s_0$.
The result of integration in Eq.\ (\ref{g1ps}) can be written as
$1 - \beta(\tau)\, z/\ell \simeq
\exp[-\beta(\tau)\, z/\ell]$ with
$\beta z/\ell \simeq z (\ell s_1)^{-1/2} = (2 s_0/s_1)^{1/2} \ll 1$.
In the opposite limit of $\beta z / \ell \gg 1$, $g_1(\vec{r}, \tau)$
vanishes and the functional form of its $z$-dependence is of no importance.
Anyway, it will be seen from the following that the exact functional form
of the $\vec{r}$-dependence of $g_1(\vec{r}, \tau)$
is not of crucial importance, since $g_1$ is
integrated over the whole medium during the calculation of
the correlation function of diffusely reflected wave.

Consider now a typical diffusely reflected wave path of length $s$.
The phase $\varphi(t)$
accumulated along  such a path is
\begin{eqnarray}
\varphi(t) = 
k_0 \int_0^s \mathrm{d} s^{\prime}\,
\left\{ n_0 + n_2 I[\vec{r}(s^{\prime}), t] \right\},
\label{phi}
\end{eqnarray}
where the integration is along the path.
The variance of the phase difference
$\Delta \varphi(\tau) =
\varphi(t+\tau) - \varphi(t)$,
averaged over all possible paths of the same length $s$,
is found directly from Eq.\ (\ref{phi}) as a sum of
two contributions:
\begin{eqnarray}
&&\left< \Delta\varphi^2(\tau) \right>_s^{(1)} =
(\pi/n_0)\, k_0 \ell\, n_2^2 \left< \Delta I(\vec{r},
\tau)^2 \right>_s s/\ell,
\label{d1} \\
&&\left< \Delta\varphi^2(\tau) \right>_s^{(2)} =
(k_0 \ell)^2 n_2^2
\left< \Delta I(\vec{r}_1, \tau)
\Delta I(\vec{r}_2, \tau) \right>_s
(s/\ell)^2,
\label{d2}
\end{eqnarray}
where 
$\Delta I(\vec{r}, \tau) = I(\vec{r}, t+\tau) - I(\vec{r}, t)$.
$\left< \Delta\varphi^2(\tau) \right>_s^{(1)}$ and
$\left< \Delta\varphi^2(\tau) \right>_s^{(2)}$
originate from the short- and the long-range spatial correlations
of $\Delta I$, respectively.

If $\tau \gg T_{jump}$, the calculation of the r.h.s.
averages in Eqs.\ (\ref{d1}) and (\ref{d2}) is straightforward.
To evaluate
$\left< \Delta I(\vec{r}, \tau)^2 \right>_s$, we integrate
$\left< \Delta I(\vec{r}, \tau)^2 \right> =
2 \left< I(\vec{r}, t) \right>^2 [1 - g_1^2(\vec{r}, \tau)]$
over $z > 0$ with a weight function $\rho_s(\vec{r})$
describing the density of paths of length $s$ in the
medium.\cite{bicout93}
To find the average of Eq.\ (\ref{d2}),
we calculate the
long-range correlation function 
$\left< \Delta I(\vec{r}_1, \tau) \Delta I(\vec{r}_2, \tau) \right>$
using the Langevin approach\cite{zyuzin87} with
the time correlation function of Langevin random sources
being proportional to $g_1^2(\vec{r}, \tau)$.
The result is then averaged over all pairs of points
$\vec{r}_1$, $\vec{r}_2$ belonging to the same path
with a weight function $\rho_s(\vec{r}_1, \vec{r}_2)$ which
is calculated similarly to $\rho_s(\vec{r})$.
As $g_1 = \exp(-\beta z/\ell)$ enters into the derivations
of $\left< \Delta I(\vec{r}, \tau)^2 \right>_s$ and
$\left< \Delta I(\vec{r}_1, \tau) \Delta I(\vec{r}_2, \tau) \right>_s$,
$\left< \Delta\varphi^2(\tau) \right>_s^{(1), (2)}$ appear to depend
explicitly on $\beta$.
The details of the calculations will be presented elsewhere,
the final results are:
\begin{eqnarray}
&&\left< \Delta\varphi^2(\tau) \right>_s^{(1)} =
2\pi\, k_0 \ell\, \Delta n^2 
H\left( \beta \sqrt{\frac{s}{3 \ell}}\; \right)
\frac{s}{\ell},
\label{d11} \\
&&\left< \Delta\varphi^2(\tau) \right>_s^{(2)} \simeq
6 \Delta n^2 \times
\cases{
\beta \times (s/\ell)^2,
&$\beta \sqrt{s/\ell} \leq 1$,
\cr
(s/\ell)^{3/2},
&$\beta \sqrt{s/\ell} > 1$,}
\label{d22}
\end{eqnarray}
where 
$H(x) = \sqrt{\pi} x \exp(x^2) [1 - \mathrm{Erf}(x)]$,
and $n_0 = 1$ is assumed for simplicity.
Applying Eq.\ (\ref{g1ps}), we obtain the autocorrelation
function of diffusely reflected wave:
\begin{eqnarray}
g_1(\ell, \tau) = \underline{\exp\left[
-\beta\left( \tau \right) \right]} =
\int_{0}^\infty \mathrm{d} s\,
P(\ell, s)
\exp\left[ -\frac{1}{2} \left< \Delta \varphi^2(\tau) \right>_s
\right]
=
\underline{F\left[ \beta\left( \tau \right) \right]},
\label{beq}
\end{eqnarray}
where
$\left< \Delta\varphi^2(\tau) \right>_s$ is a sum of the terms
given by Eqs.\ (\ref{d11}) and (\ref{d22}).
Since the integral of Eq.\ (\ref{beq}) does not contain $\tau$ in
explicit form, we conclude that $\beta$ and $g_1$ are independent
of $\tau$, as expected for $\tau \gg T_{jump}$. From here on, $g_1$
without arguments denotes the value of the autocorrelation function of
diffusely reflected wave at $\tau \gg T_{jump}$.

In Eq.\ (\ref{beq}), we have introduced a new function
$F(\beta)$ which depends on $\Delta n$ and $k_0 \ell$ as well.
The underlined part of Eq.\ (\ref{beq}) can be considered as
a nonlinear equation for $\beta$,
or, equivalently, for $g_1$ since $\beta = -\ln g_1$ by definition.
Making a simultaneous plot of $F(\beta)$ and
$\exp(-\beta)$ (see the lower solid curve
in the inset of Fig.\ \ref{fig1}),
we find that Eq.\ (\ref{beq}) has two solutions:
$\beta = 0$ and $\beta > 1$.
The first solution is ``unstable'' with respect to the
weak noise $\delta J$ in Eq.\ (\ref{weq}), since $\delta J \neq 0$
introduces an additional dephasing term into Eq.\ (\ref{beq}), yielding
$F(0) < 1$. It is then the
second solution $\beta > 1$ which is the physically realizable one.
As $\beta > 0$ implies $g_1 < 1$, our results show that
$\psi(\vec{r}, t)$ exhibits
{\em spontaneous\/} fluctuations, and that the speckle
pattern resulting from the multiple scattering of coherent
wave in a weakly nonlinear disordered medium is {\em unstable.}

The value of $g_1$, characterizing the magnitude of
spontaneous fluctuations ($g_1 = 1$ in the absence of
fluctuations), can be found by a numerical solution
of Eq.\ (\ref{beq}) (see the dashed line in Fig.\ \ref{fig1}),
as well as analytically. Assuming $1 - g_1 \ll 1$,
and noting that the ultimate contribution to
$\left< \Delta\varphi^2(\tau) \right>_s$ is given either
by Eq.\ (\ref{d11}) [for $s/\ell \ll (k_0 \ell)^2$] or by
Eq.\ (\ref{d22}) [for $s/\ell \gg (k_0 \ell)^2$], we obtain
\begin{eqnarray}
g_1 \simeq 1-\cases{
2 \left| \Delta n \right|^{2/3},
&$\left| \Delta n \right| < (k_0 \ell)^{-3/2}$, \cr
3 \left| \Delta n \right| \sqrt{k_0 \ell},
&$\left| \Delta n \right| > (k_0 \ell)^{-3/2}$.}
\label{b0}
\end{eqnarray}

Up to now, we have considered a nonabsorbing medium.
Absorption, however, is inevitable in a real medium and
should therefore be necessarily taken into account. This can be done
by allowing $\varepsilon_0$ in Eq.\ (\ref{weq}) to have a small
imaginary part. Eq.\ (\ref{d11}) then becomes
\begin{eqnarray}
\left< \Delta\varphi^2(\tau) \right>_s^{(1)} =
2 \pi\, k_0 \ell\, \Delta n^2 \left\{ 
H\left[ \left(\beta + \frac{\ell}{L_a} \right)
\sqrt{\frac{s}{3 \ell}}\; \right] -
H\left[ \frac{\ell}{L_a}
\sqrt{\frac{s}{3 \ell}}\; \right] \right\}
\frac{s}{\ell},
\label{d111}
\end{eqnarray}
where $L_a = [\ell/(3 k_0 \mathrm{Im}\, \varepsilon_0)]^{1/2}
\gg \ell$ is the absorption length,
and $n_0^2 = \mathrm{Re}\, \varepsilon_0 = 1$.
Eq.\ (\ref{d22}) is still approximately valid for
$(\ell/L_a) (s/\ell)^{1/2} \leq 1$, while for
$(\ell/L_a) (s/\ell)^{1/2} > 1$ we get
\begin{eqnarray}
\left< \Delta\varphi^2(\tau) \right>_s^{(2)} \simeq
6 \Delta n^2 \times
\cases{
\beta \times (L_a/\ell)^3 \sqrt{s/\ell},
&$\beta \sqrt{s/\ell} \leq 1$,
\cr
(L_a/\ell)^3,
&$\beta \sqrt{s/\ell} > 1$.}
\label{d222}
\end{eqnarray}

Graphical solution of Eq.\ (\ref{beq}) with
$P(\ell, s) \propto s^{-3/2} \exp[-3\ell/(4s) - s\ell/(3 L_a^2)]$
is presented in the inset of Fig.\ \ref{fig1}
for several absorption lengths $L_a$ . 
As follows from the figure, the absorption has to be weak for
the second solution of Eq.\ (\ref{beq}), $\beta > 0$, to exist.
Sufficiently strong absorption prevents spontaneous
fluctuations of $\psi(\vec{r}, t)$, stabilizing the
speckle pattern. The condition for the onset of the speckle
pattern instability is 
$(\partial/\partial \beta) F(\beta) = -1$
evaluated at $\beta = 0$. This yields:
\begin{eqnarray}
p =
\Delta n^2 \left(\frac{L_a}{\ell} \right)^2
\left[ k_0 \ell + \frac{L_a}{\ell} \right] \simeq 1,
\label{stbl}
\end{eqnarray}
where we introduce a control parameter $p$.
If $p<1$, $g_1 = 1$ and the
multiple-scattering speckle pattern is stable,
while for $p>1$, an instability shows up leading to
$g_1 < 1$ (see Fig.\ \ref{fig1}).
While $g_1$ characterizes the mutual resemblance of solutions
of Eq.\ (\ref{weq}), the number of solutions grows exponentially
with $p^{3/4}$, as can be estimated using the method of
Ref.\ \onlinecite{spivak00}.
The striking feature of Eq.\ (\ref{stbl}) is that
$p$ can reach unity even for very small $\left| \Delta n \right|$
provided that the extensive parameter $L_a/\ell$ is large enough.
The stability condition $p < 1$ is therefore never satisfied in
a nonabsorbing medium.
Eq.\ (\ref{stbl}) is
consistent with the result of Spivak and Zyuzin,\cite{spivak00}
who have shown that the perturbation analysis
of the sensitivity of speckle
pattern in a nonlinear disordered medium
to changes of scattering potential
fails for $\Delta n^2 (L/\ell)^3 > 1$, where $L$ is
the typical size of the medium.

The physical origin of the instability of speckle pattern for $p>1$
can be revealed by realizing that the system ``coherent wave + nonlinear
disordered medium'' has a positive distributed feedback.
An infinitely weak perturbation of the wave intensity $I(\vec{r}, t)$
produces a change
of the local refractive index, altering the phases of waves
propagating in the medium, and affecting their
mutual interference. Since it is this interference which determines
$I(\vec{r}, t)$, the feedback loop is built up and closed.
For $p \gtrsim 1$, the feedback is sufficiently strong
to compensate for the (diffusive on average) spreading of the
initial intensity perturbation, and the speckle pattern $I(\vec{r}, t)$
becomes unstable.

The above analysis can be generalized to the case of
$\delta \varepsilon$ which varies slowly with time (moving scatterers).
An additional contribution to $\left< \Delta\varphi^2(\tau) \right>_s$
arises in this case.
If, for instance, the scatterers undergo Brownian
motion with a diffusion coefficient $D$, this contribution is
\begin{eqnarray}
\left< \Delta\varphi^2(\tau) \right>_s^{(0)} =
\frac{\tau}{\tau_0} \left\{
1 + 2 \Delta n \left[
1 - H \left( \frac{\ell}{L_a} \sqrt{\frac{s}{12 \ell}}\; \right)
\right] \right\} \frac{s}{\ell},
\label{d0}
\end{eqnarray}
where both $\tau_0 = (4 k_0^2 D)^{-1}$ and $\tau$ are
assumed to be much larger than $T_{jump}$. The time autocorrelation
functions of diffusely reflected waves obtained by solving
Eq.\ (\ref{beq}) with $\left< \Delta\varphi^2(\tau) \right>_s$
given by a sum of contributions of Eqs.\ (\ref{d22}),
(\ref{d111}), (\ref{d222}),
and (\ref{d0}) are shown in Fig.\ \ref{fig2}.
The two lower curves correspond to the unstable regime ($p > 1$),
and therefore $g_1(\tau) < 1$ even if $\tau/\tau_0 = 0$.
For the upper curve, the speckle pattern is stable ($p < 1$) and
$g_1(\tau) = 1$ at $\tau/\tau_0 = 0$.

In conclusion, we have shown that the multiple-scattering speckle
pattern may become unstable if the disordered medium exhibits a
weak nonlinearity of Kerr type.
The instability is due to the distributed feedback provided by
the scattering, and its development is independent of
the sign of nonlinearity.
The instability threshold decreases
as absorption is reduced, and the speckle pattern is never
stable in a nonabsorbing unbounded medium.
Finally, we note that our results may be of particular
importance in the context of ``random laser'',\cite{letokhov68}
where absorption may be efficiently compensated by stimulated emission.

\begin{figure}
\caption{Main plot: a ``bifurcation diagram'' of the speckle pattern
resulting from the multiple scattering of a coherent wave in
a semi-infinite nonlinear disordered medium at fixed $k_0 \ell = 100$.
The inverse absorption lengths $\ell/L_a$ are indicated
near each curve, dashed line corresponds to a nonabsorbing medium,
dotted line shows $g_1 \equiv 1$ (linear case).
In the absence of absorption,
the speckle pattern is unstable ($g_1 < 1$) at any
$\Delta n \neq 0$, while in a dissipative medium,
$\left| \Delta n \right|$
should exceed a certain threshold value for
the instability to develop.
The threshold values of $\left| \Delta n \right|$
following from Eq.\ (\ref{stbl}) are shown by arrows.
Inset: the graphical solution of Eq.\ (\ref{beq})
for fixed $\left| \Delta n \right|$, $k_0 \ell$, and the values of 
$\ell/L_a$ indicated near each curve.
The dash-dotted line is $\exp(-\beta)$.}
\label{fig1}
\end{figure}

\begin{figure}
\caption{The field autocorrelation function
of a wave diffusely reflected from a
semi-infinite nonlinear (solid curves) or linear (dotted curves)
disordered medium for fixed $\Delta n$, $k_0 \ell$, and inverse
absorption lengths $\ell/L_a$ indicated near each curve
($T_{jump}/\tau_0 \rightarrow 0$ is assumed).
For the two largest values
of absorption length, the ``nonlinear'' speckle pattern is
unstable ($g_1 < 1$ at $\tau/\tau_0 = 0$).}
\label{fig2}
\end{figure}

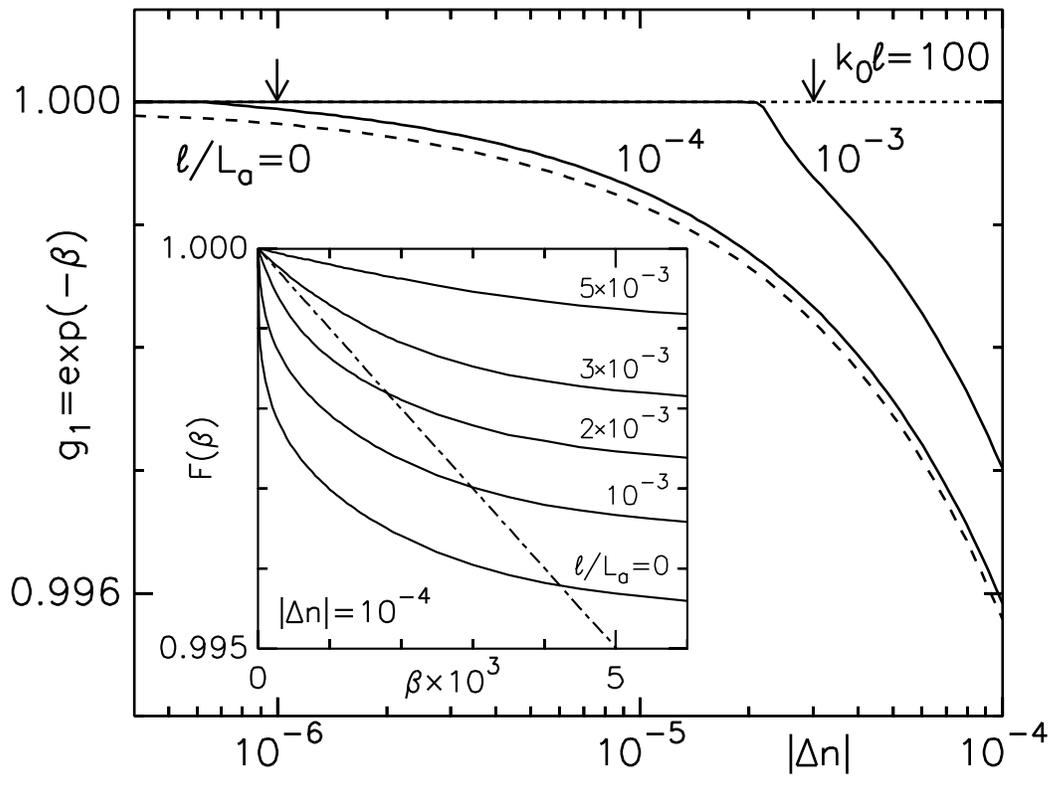
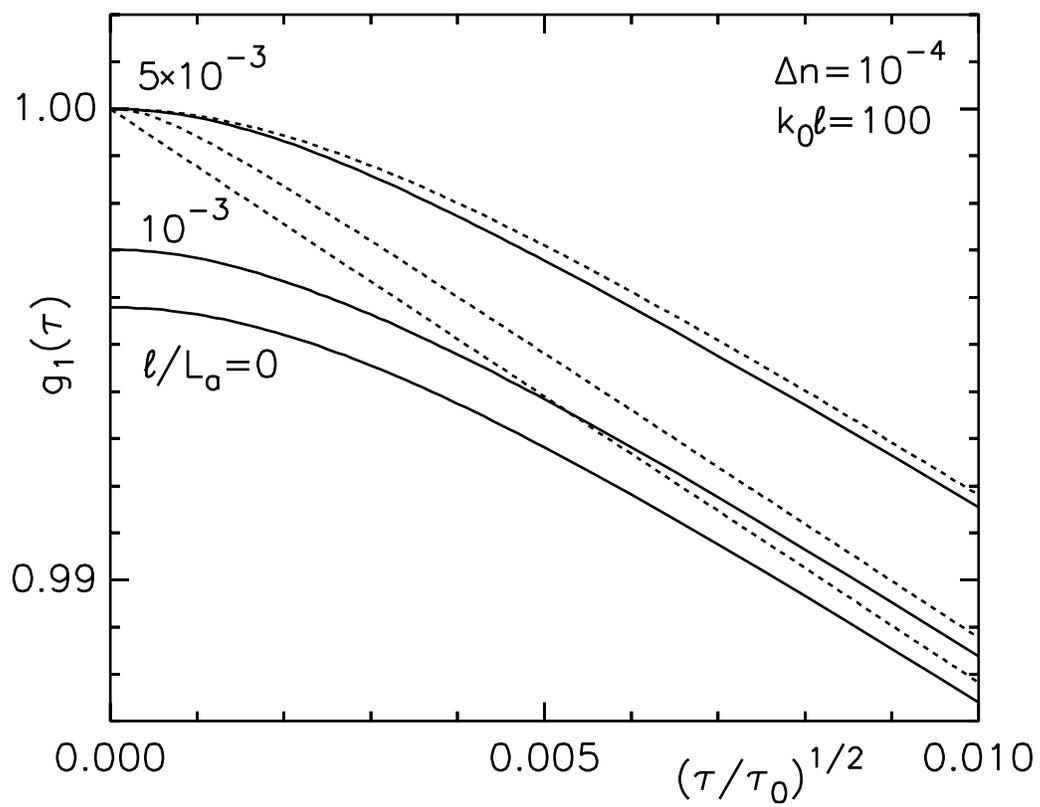
\end{document}